\documentclass[aps,pra, showpacs, superscriptaddress, onecolumn,
preprint]{revtex4}

\usepackage{graphicx}

\newcommand{\Xh}{\ensuremath{\chi_{\text{hom}}}}
\newcommand{\Xl}{\ensuremath{\chi_{\text{line}}}}
\newcommand{\Xt}{\ensuremath{\chi_{\text{tot}}}}

\newcommand{\subs}[1]{{\mbox{\tiny #1}}}
\newcommand{\Velec}{\ensuremath{v_\subs{el}}}

\begin{document}

\title{Quantum key distribution over 25 km with an all-fiber
continuous-variable system}

\author{J\'er\^ome Lodewyck}
\affiliation{Thales Research and Technologies, RD 128,
91767 Palaiseau Cedex, France}
\affiliation{Laboratoire Charles Fabry de l'Institut d'Optique --- CNRS ---
Univ. Paris-Sud, Campus
Polytechnique, RD 128, 91127 Palaiseau Cedex, France}

\author{Matthieu Bloch}
\affiliation{GeorgiaTech --- CNRS UMI 2958, 2-3 rue Marconi, 57070 Metz,
France}

\author{Ra\'ul Garc\'ia-Patr\'on}
\affiliation{Centre for Quantum Information and Communication, \'Ecole
Polytechnique, CP 165/59, Universit\'e libre de Bruxelles, 
50 av. F. D. Roosevelt, B-1050 Brussels, Belgium}

\author{Simon Fossier}
\affiliation{Thales Research and Technologies, RD 128,
91767 Palaiseau Cedex, France}
\affiliation{Laboratoire Charles Fabry de l'Institut d'Optique --- CNRS ---
Univ.
Paris-Sud, Campus
Polytechnique, RD 128, 91127 Palaiseau Cedex, France}

\author{Evgueni Karpov}
\affiliation{Centre for Quantum Information and Communication, \'Ecole
Polytechnique, CP 165/59, Universit\'e libre de Bruxelles, 
50 av. F. D. Roosevelt, B-1050 Brussels, Belgium}

\author{Eleni Diamanti}
\affiliation{Laboratoire Charles Fabry de l'Institut d'Optique --- CNRS ---
Univ.
Paris-Sud, Campus
Polytechnique, RD 128, 91127 Palaiseau Cedex, France}

\author{Thierry Debuisschert}
\affiliation{Thales Research and Technologies, RD 128,
91767 Palaiseau Cedex, France}

\author{Nicolas J. Cerf}
\affiliation{Centre for Quantum Information and Communication, \'Ecole
Polytechnique, CP 165/59, Universit\'e libre de Bruxelles, 
50 av. F. D. Roosevelt, B-1050 Brussels, Belgium}

\author{Rosa Tualle-Brouri}
\affiliation{Laboratoire Charles Fabry de l'Institut d'Optique --- CNRS ---
Univ.
Paris-Sud, Campus
Polytechnique, RD 128, 91127 Palaiseau Cedex, France}

\author{Steven W. McLaughlin}
\affiliation{GeorgiaTech --- CNRS UMI 2958, 2-3 rue Marconi, 57070 Metz,
France}

\author{Philippe Grangier}
\affiliation{Laboratoire Charles Fabry de l'Institut d'Optique --- CNRS ---
Univ.
Paris-Sud, Campus
Polytechnique, RD 128, 91127 Palaiseau Cedex, France}

\begin{abstract}
We report on the implementation of a reverse-reconciliated coherent-state
continuous-variable quantum key distribution system, with which we generated
secret keys at a rate of more than 2 kb/s over 25 km of optical fiber. Time
multiplexing
is used to transmit both the signal and phase reference in the same optical
fiber. Our system includes all experimental aspects required for a field
implementation of a quantum key distribution setup. Real-time reverse
reconciliation is achieved by using fast and efficient LDPC error correcting
codes.
\end{abstract}

\pacs{03.67.Dd, 42.50.Lc, 42.81.-i, 03.67.Mn}

\maketitle

\section{Introduction}
\label{sec:intro}

Quantum Key Distribution (QKD) enables two remote parties, Alice and Bob, linked
by a quantum channel and an authenticated classical channel, to share a common
random binary key that is unknown to a potential eavesdropper, Eve. Many
QKD protocols~\cite{gisin} encode key information in discrete variables
of
single photon light pulses, such as polarization or
phase~\cite{bennett:BB84, bennett:phase, bennett:entanglement, ekert,
franson,
yamamoto, buttler, cow}.
Recently, other protocols using so-called continuous variables (CV), such as
both quadratures of a coherent state, have been
proposed~\cite{ralph,hillery,cerf:squeezeQKD,grosshans:prl02,grosshans:nature,
postselection}.
Channel symbols are formed by conjugate continuous quantum variables,
linked by Heisenberg inequalities. The secrecy of the QKD protocol
is based on the resulting quantum uncertainty relations. Such protocols
eliminate the
need for single photon technology, as they only require standard off-the-shelf
telecom components --- such as diode lasers, electro-optics modulators, and PIN
photodiodes --- compatible with high repetition rates. On the other hand, CVQKD
protocols require elaborate classical error correction algorithms to efficiently
extract secret bits from correlated continuous variables.

In this paper, we describe a complete implementation of the coherent-state
reverse-reconciliated (RR) CVQKD protocol described in~\cite{grosshans:nature}.
In this protocol, the quadratures $x$ and $p$ of a train of coherent-state
pulses are modulated in the complex plane with a centered bi-variate Gaussian
modulation
of variance $V_A N_0$, where $N_0$ is the shot noise variance that appears in
the Heisenberg relation $\Delta x \Delta p \geq N_0$. These coherent states are
sent from Alice to Bob through the quantum channel, along with a strong phase
reference --- or local oscillator (LO). Upon reception, Bob randomly
measures the $x$ or $p$ quadrature by making the signal interfere
with the LO
in a
pulsed, shot-noise limited homodyne detector. This protocol allows Alice and Bob
to share a set of correlated Gaussian data. A random fraction of this set
is
publicly revealed to probe the transmission parameters, while the remaining part
is used to build a secret key based on Bob's data. This is achieved in practice
with a classical error correction scheme called ``Multi-Level Coding" using
efficient one-way Low Density Parity Check (LDPC) codes. We report the full
implementation of both quantum and classical parts of this RRCVQKD protocol over
a standard single-mode telecom fiber of 25~km, leading to a final secret key
distribution rate of more than 2 kb/s.

\section{Theoretical evaluation of the secret key rates}
\label{sec:rates}

In this Section, we detail the calculation of the secret key rates that are
available to Alice and Bob when applying the RRCVQKD
protocol. In QKD, one
evaluates the secret key rate by upper bounding the information that the
adversary, Eve, can acquire in the worst case. This is typically done under the
following assumptions: (i) Eve has no limit in terms of computational power;
(ii) Eve has full control over the quantum channel, and is only limited in her
action on this channel by the laws of quantum physics; (iii) Eve can freely
monitor the classical public channel used for key distillation, but she cannot
modify the messages (authenticated channel); (iv) Eve has no access to the
laboratories (apparatuses) of Alice and Bob. Traditionally, the type of attacks
that Eve can implement are ranked by increasing power into three classes,
depending on how exactly she interacts with the pulses sent by Alice with
auxiliary pulses (ancillae), and on when she measures these ancillae.
 The theoretical bound on Eve's information depends on the
class of attacks that is considered:
\begin{itemize}
	\item \emph{Individual} attack: Eve interacts individually with each
pulse
sent by Alice, and stores her ancilla in a quantum memory. She then performs an
appropriate measurement on her ancilla after the sifting procedure (during
which Bob
reveals whether he chose to measure $x$ or $p$), but before the
key
distillation stage (in particular, before error correction). Using this
attack, the
maximum information accessible to Eve is
bounded by the classical (Shannon~\cite{shannon:1948, shannon:1949}) mutual
information $I_{BE}$ on
Bob's data. Moreover, in the case of continuous-variable QKD, it is known that
the optimal individual attack is a Gaussian operation~\cite{grosshans:prl04},
which considerably restricts the set of attacks that need to be considered and
yields a simple closed formula for $I_{BE}$.
	\item \emph{Collective} attack: Eve interacts individually with each
pulse
sent by Alice but, instead of measuring immediately after sifting, she listens
to the communication between Alice and Bob during the key distillation
procedure, and only then applies the optimal collective measurement on the
ensemble of stored ancillae. In this attack, the maximum information she may
have access to is limited by the Holevo bound $\chi_{BE}$~\cite{holevo:canal}.
As in the case of
individual attacks against continuous-variable QKD, Gaussian attacks have been
shown to be optimal among all collective
attacks~\cite{garcia-patron:prl,navasques:prl06}, which results in a simple
expression for $\chi_{BE}$.
	\item \emph{Coherent} attack: This is the most powerful attack that
Eve can
implement. Here, she is allowed to interact collectively with all the pulses
sent by
Alice, and, after having monitored the key distillation messages, she applies an
optimal joint measurement over all the ancillae. The security with respect to
this kind of attacks is more complicated to address, but, under the assumption
of the symmetry of the privacy amplification and channel probing protocols, it
was proven for discrete-variable QKD in~\cite{R05} (and conjectured for
continuous-variable QKD in~\cite{garcia-patron:prl,navasques:prl06}) that
coherent attacks are not more efficient than collective attacks. This step is
quite important as it ensures unconditional security as long as one has a
security proof with respect to collective attacks, for which the key rates are
far simpler to evaluate.\end{itemize}

In the following, we will consider individual and collective attacks, for which
the security analysis lies on firm grounds. We will then derive expressions for
$I_{BE}$ and $\chi_{BE}$ as a function of the losses and of the noise of the
quantum
channel, assuming as usual that Eve can take both of them to her advantage.
We will restrict our study to Gaussian attacks, which have been shown to be
optimal~\cite{garcia-patron:prl,navasques:prl06}; this significantly simplifies
the
calculation of the secret key rates since we only have to consider covariance
matrices. It is known
that Alice and Bob can distill perfectly correlated secret key bits provided
that the amount of information they share, $I_{AB}$, remains higher than the
information acquired by Eve ($I_{BE}$ or
$\chi_{BE}$ for reverse reconciliation). In this strictly
information-theoretic point of view, and in the case of RR,
we define the ``raw'' key rate as $\Delta
I^{\text{Shannon}}
= I_{AB} - I_{BE}$, or respectively $\Delta I^{\text{Holevo}} = I_{AB} -
\chi_{BE}$.

\subsection{Entanglement-based CVQKD scheme}

An usual prepare-and-measure (P\&M) implementation of a Gaussian protocol with
coherent states has been described in Section~\ref{sec:intro}, and consists in a
quantum transmission followed by a classical data processing. During the quantum
part,
Alice randomly generates two numbers $(x_A,p_A)$ from a Gaussian distribution,
prepares a coherent state centered on $(x_A,p_A)$, and sends it to Bob through
the quantum channel. Bob receives this state, and randomly measures
the quadrature $x$ or $p$ by choosing the appropriate phase for his
homodyne measurement.

\begin{center}
\begin{figure*}[tb]
	\includegraphics[width=.8 \textwidth]{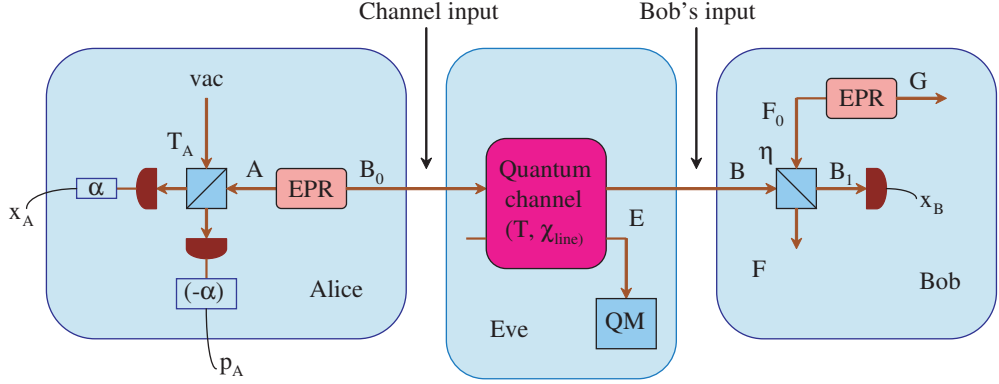}
	\caption{(Color online) Entanglement-based scheme of CVQKD.
The transmittance $T_A$ and $\eta$ characterize the measurements
at Alice's and Bob's sides, while the channel transmittance $T$ and added noise
$\Xl$ are controlled by Eve. The QM box corresponds to Eve's quantum memory.}
	\label{fig:EPR}
\end{figure*}
\end{center}

As defined in Fig.~\ref{fig:EPR}, the quantum channel is characterized by its
transmission $T\le 1$ and its excess noise
$\varepsilon$ such that the noise variance at Bob's input is
$(1+T\varepsilon)N_0$. We call $\Xl=1/T-1+\varepsilon$ the total channel added
noise referred to the channel input, which is
composed of the noise due to losses
$1/T-1$ and the excess noise $\varepsilon$. With these notations, all noises are
expressed in shot noise units. The signal then reaches Bob's detector, which is
modeled by assuming that the signal is further attenuated by a factor $\eta$
(detection losses) and mixed with some thermal noise (electronic noise $\Velec$
added by the detection electronics, expressed in shot noise units). The total
noise introduced by the realistic homodyne detector is $\Xh=(1+\Velec)/\eta-1$,
when referred to Bob's input. The total noise added between Alice and Bob
then reads $\Xt=\Xl+\Xh/T$, referred to the channel input.

In the following, we will exploit the fact that this P\&M description of
QKD with Gaussian states is equivalent to the entanglement-based
(EB) scheme presented in Fig.~\ref{fig:EPR}, which simplifies the theoretical
calculation of the key rates and
provides a unified description of the different existing
protocols~\cite{grosshans:qic}. The main idea is to
view Alice's quantum state preparation as resulting from the measurement of one
half of a two-mode squeezed vacuum state (EPR state). The second half of the EPR
state corresponds to the state sent to Bob through the quantum channel. The
Gaussian state $AB_0$ is completely determined by its covariance matrix
$\gamma_{AB_0}$, which has the form
\begin{eqnarray}
	\label{EPR}
	&\gamma^{\rm EPR}_V
	= \left[
	\begin{array}{cc}
		V \cdot \openone & \sqrt{V^2-1}\cdot \sigma_z \\
		\sqrt{V^2-1} \cdot \sigma_z & V \cdot \openone
	\end{array}
	\right]&\\
	&{\rm where}\quad
	\openone=\left[\begin{array}{cc}1&0\\0&1\end{array}\right]
	\quad {\rm and} \quad
	\sigma_z=\left[\begin{array}{cc}1&0\\0&-1\end{array}\right]&\nonumber
\end{eqnarray}
where $V$ is the variance, in shot noise units, of the thermal state that we
observe if we trace
out $A$. This thermal state also corresponds exactly to the thermal state
observed at the output of Alice's station if we implement a P\&M protocol,
resulting from the ensemble of Gaussian-modulated coherent states (with some
specific Gaussian
distribution)~\cite{cerf:squeezeQKD,grosshans:prl02,grosshans:nature,R04}. In
fact, every
P\&M scheme can be rigorously translated into an EB scheme. First, the generated
states in a P\&M scheme are equivalent to the states on which mode $B_0$ is
projected after Alice's measurement in an EB scheme. Second, the modulation of
the states in a P\&M scheme corresponds in the EB scheme to the variation of the
mean value of the state of mode $B_0$ conditioned on Alice's measurement. This
implies that the modulation in the P\&M scheme is directly related to Alice's
measurement in the EB scheme via a one-to-one correspondence.

As an example, Alice applying a homodyne detection of $x_A$ ($T_A=1$)
corresponds to projecting the mode $B_0$ onto squeezed states that are
displaced
according to a Gaussian distribution of the measured quadrature $x_A$. This is
exactly equivalent to the protocol proposed in~\cite{cerf:squeezeQKD}. If she
applies instead a heterodyne measurement ($T_A=1/2$), she prepares coherent
states modulated over a bi-dimensional Gaussian distribution of variance
$V_AN_0$, as in~\cite{grosshans:nature,grosshans:prl02}. Let us focus on the
equivalence between the EB scheme and the P\&M scheme in this case. In the P\&M
scheme, Alice randomly chooses the values $x_A$ and $p_A$ distributed according
to a Gaussian distribution centered on zero and of variance $V_AN_0$, and sends
Bob a coherent state ($V_{B_0|A}=1$ in shot noise units) centered on
($x_A,p_A$). In the EB scheme, Alice estimates the quadratures $x_{B_0}$ and
$p_{B_0}$ of the state sent to Bob by multiplying the outcomes of her
measurements by a factor $\alpha=\sqrt{2\frac{V-1}{V+1}}$ (with a minus sign for
$p$-quadrature)~\cite{grosshans:qic}. Her uncertainty on the inferred values of
$x_{B_0}$ and $p_{B_0}$ for a given $x_A$ and $p_A$ is exactly $V_{B_0|A}=1$,
which corresponds to the uncertainty of a coherent state in the P\&M scheme. The
inferred values of $x_{B_0}$ and $p_{B_0}$ are distributed according to a
Gaussian distribution of variance $V_AN_0 = (V-1)N_0$, which coincides with
Alice's modulation in the P\&M scheme.

Note that the EB scheme allows us, at the same time, to simplify the description
of the realistic detector at Bob side. As shown in Fig. 1, the inefficiency of
Bob's detector is modeled by a beam splitter with transmission $\eta$, while the
electronic noise $\Velec$ of Bob's detector is modeled by a thermal state
$\rho_{F_0}$ with variance $V_N N_0$ entering the other input port of the beam
splitter, so that $V_N = 1 + \Velec/(1-\eta)$. Considering the thermal state
$\rho_{F_0}$ as the reduced state obtained from a two-mode squeezed state
$\rho_{F_0G}$ of variance $V_N N_0$ allows us to simplify the calculations.

\subsection{Individual attack --- Shannon rate}

The mutual information $I_{AB}$ is calculated directly from the variance $V_B
N_0$ of the quadratures measured by Bob, with $V_B=\eta T\left(V+\Xt\right)$,
and the conditional variance $V_{B|A}=\eta T(1+\Xt)$ using Shannon's equation
\begin{equation}
	\label{IAB}
I_{AB}=\frac{1}{2}\log_2\frac{V_B}{V_{B|A}}=\frac{1}{2}\log_2\frac{V+\Xt}
 {
1+\Xt}.
\end{equation}
In an individual attack, Eve performs her measurements just after Bob reveals
the quadrature he has measured (sifting) but before the error correction. Her
information is thus restricted to the Shannon information accessible in her
ancilla after measurement, and is bounded using the entropic uncertainty
relations as proven in~\cite{grosshans:prl04}. In the RR protocol, the reference
during the error correction protocol being Bob, Eve's information reads
\begin{eqnarray}
	I_{BE}&=&\frac{1}{2}\log_2\frac{V_B}{V_{B|E}}\\
	\textrm{where}\quad V_B=\eta T(V+\Xt)	\quad &\textrm{and}&\quad
V_{B|E}
=
\eta\left[\frac{1}{T(1/V+\Xl)}+\Xh\right].\nonumber
\end{eqnarray}
Note that we have considered the so-called ``realistic
model'' suggested
in~\cite{grosshans:nature}, where Eve cannot benefit from the noise added by
Bob's apparatus, $\Xh$. The Shannon ``raw'' key rate, proven secure against
Gaussian or non-Gaussian, individual or finite-size
attacks~\cite{grosshans:prl04}, then reads $\Delta I^{\text{Shannon}} =
I_{AB}-I_{BE}$.

\subsection{Collective attack --- Holevo rate}

In this case, the mutual information between Alice and Bob remains the same as
in the case of individual attacks, namely Eq.~(\ref{IAB}). However, Eve's
accessible information is now upper bounded by the Holevo quantity~\cite{R05},
\begin{equation}
	\label{Ichi}
	\chi_{BE} = S(\rho_E)-\int{\rm d}x_B \; p(x_B) \; S(\rho_E^{x_B}),
\end{equation}
where $p(x_B)$ is the probability distribution of Bob's measurement outcomes,
$\rho_E^{x_B}$ is the state of Eve's system conditional on Bob's measurement
outcome $x_B$, and $S(\rho)$ is the von Neumann entropy of the quantum state
$\rho$~\cite{neumann}. For an $n$-mode Gaussian state $\rho$, this entropy reads
\begin{equation}
	\label{vNe}
	S(\rho)=\sum_i G\left(\frac{\lambda_i-1}{2}\right),
\end{equation}
where $ G(x)=(x+1)\log_2(x+1)-x\log_2x$ and $\lambda_i$ are the symplectic
eigenvalues of the covariance matrix $\gamma$ characterizing $\rho$. The
calculation of Eve's information $\chi_{BE}$ is done using the following
technique. First, we use the fact that Eve's system $E$ purifies $AB$, so that
$S(\rho_E)=S(\rho_{AB})$. Second, after Bob's projective measurement resulting
in $x_B$, the system $AEFG$ (see~Fig.~1) is pure, so that
$S(\rho_E^{x_B})=S(\rho_{AFG}^{x_B})$, where $S(\rho_{AFG}^{x_B})$ is
independent of $x_B$ for protocols with Gaussian modulation of Gaussian states.
Thus, Eq.~(\ref{Ichi}) becomes
\begin{equation}
	\label{Ichi2}
	\chi_{BE}=S(\rho_{AB})-S(\rho_{AFG}^{x_B}),
\end{equation}
and can be calculated from the covariance matrix $\gamma_{AB}$ that is inferred
from the channel probing, the detector efficiency $\eta$, and the detector
electronic noise $\Velec$.

The entropy $S(\rho_{AB})$ is calculated from the symplectic eigenvalues
$\lambda_{1,2}$ of the covariance matrix
\begin{eqnarray}
	\label{gab}
	\gamma_{AB} &=& \left[
	\begin{array}{cc}
		\gamma_{A} & \sigma_{AB} \\
		\sigma^T_{AB} & \gamma_{B}
	\end{array}
	\right]\\
	&=& \left[
	\begin{array}{cc}
		V \cdot \openone & \sqrt{T(V^2-1)}\cdot \sigma_z\\
		\sqrt{T(V^2-1)}\cdot \sigma_z & T(V+\Xl)\cdot \openone
	\end{array}
	\right]\nonumber
\end{eqnarray}
The symplectic eigenvalues of $\gamma_{AB}$ are given by
\begin{equation}
	\label{L12}
	\lambda^{2}_{1,2}=\frac{1}{2}\left[A \pm \sqrt{A^2-4B}\right],
\end{equation}
where $A  =  V^2(1-2T)+2T+T^2(V+\Xl)^2$ and $B  =  T^2 (V\Xl+1)^2$.
Similarly, the entropy $S(\rho^{x_B}_{AFG})$ is determined from the symplectic
eigenvalues $\lambda_{3,4,5}$ of the covariance matrix characterizing the state
$\rho^{x_B}_{AFG}$ after Bob's projective measurement, namely
\begin{equation}
	\label{gammaxb}
	\gamma^{x_B}_{AFG}= \gamma_{AFG}-\sigma^T_{AFG;B_1}(X\gamma_BX)^\mathrm{
MP}\sigma_{AFG;B_1},
\end{equation}
where $X=\left[\begin{array}{cc}1&0\\0&0\end{array}\right]$ and MP stands for
the Moore Penrose inverse of a matrix. The matrices $\sigma_{AFG;B_1}$ in Eq.
(\ref{gammaxb}) can be read in the decomposition of the matrix
\begin{eqnarray*}
		\gamma_{AFGB_1}=\left[\begin{array}{cc}
			\gamma_{AFG} & \sigma^T_{AFG;B_1}\\
			\sigma_{AFG;B_1} & \gamma_{B_1}
			\end{array}
		\right]
\end{eqnarray*}
which is obtained by rearranging the lines and columns of the matrix describing
the system $AB_1FG$ (see Fig. 1),
\begin{eqnarray}
	\label{gammaBS}
	\gamma_{AB_1FG}
	= 	Y^T 	
\left[\gamma_{AB}\oplus\gamma^{EPR}_{F_0G}\right] Y\\
	{\rm where~~} Y = \left(\openone_{A}\oplus
S^{BS}_{BF_0}\oplus\openone_G\right).\nonumber
\end{eqnarray}
This matrix is obtained by applying onto systems $B$ and $F_0$ a beam splitter
transformation ($S^{BS}_{BF_0}$) that models the efficiency $\eta$ of Bob's
detector, where $F_0$ is the thermal state that models the electronic noise of
the detector $\Velec$. A long but straightforward calculation
shows that the symplectic eigenvalues $\lambda_{3,4}$ are given by
\begin{eqnarray}
	\label{L34}
	\lambda^2_{3,4}&=& \frac{1}{2}(C\pm\sqrt{C^2-4D})\\
	{\rm where~~}
	\label{AF1}
	C  &=&  \frac{V\sqrt{B}+ T(V+\Xl) + A \Xh}{T(V+\Xt)}\nonumber\\
	 {\rm and} \quad
	D  &=&  \sqrt{B}\frac{V + \sqrt{B} \Xh }{T(V+\Xt)}.
	\label{BF1}\nonumber
\end{eqnarray}
while the last symplectic eigenvalue is simply $\lambda_5=1$.

The Holevo information bound then reads
\begin{eqnarray}
	\chi_{BE} = &G&\left(\frac{\lambda_1-1}{2}\right) +
	G\left(\frac{\lambda_2-1}{2}\right)\\
	 - &G&\left(\frac{\lambda_3-1}{2}\right) -
	G\left(\frac{\lambda_4-1}{2}\right)\nonumber
\end{eqnarray}
and the Holevo ``raw'' key rate, proven secure against collective attacks, reads
$\Delta I^{\text{Holevo}} = I_{AB} - \chi_{BE}$.

\section{Implementation of continuous-variable quantum key distribution}
\label{sec:implementation}

\subsection{Experimental setup}
\label{sec:setup}

\begin{figure*}[tb]
	\includegraphics[width=.8\textwidth]{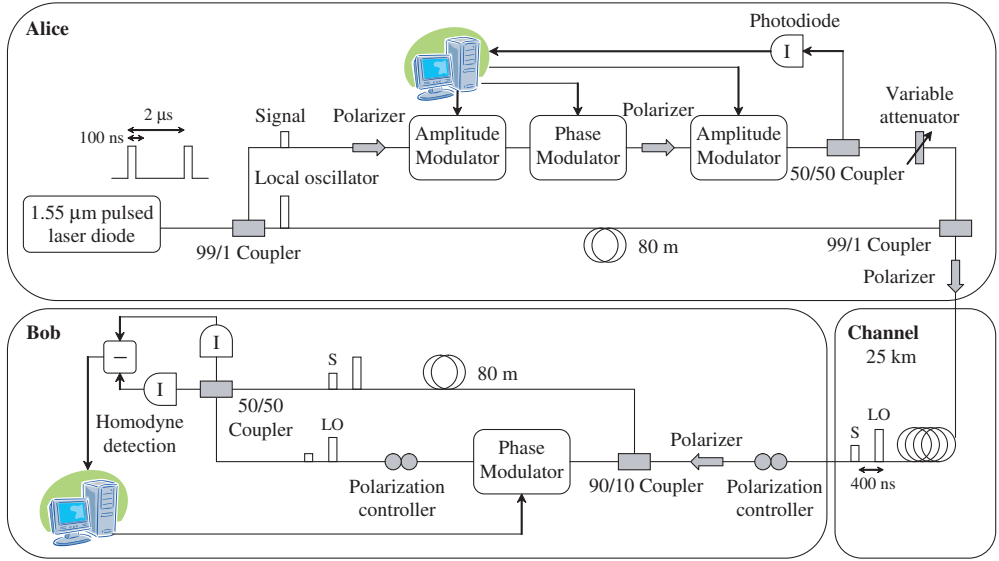}
	\caption{(Color online) Experimental setup for CVQKD.}
	\label{fig:setup}
\end{figure*}

The experimental setup for the CVQKD
experiments that we have performed is shown in Fig.~\ref{fig:setup}. It is a
coherent-state QKD setup, operating at 1550~nm and consisting entirely of
standard fiber optics and telecommunication components. Alice uses a laser
diode, pulsed with a repetition rate of 500~kHz, to generate pulses with a width
of 100~ns. Using a highly asymmetric fiber-optic coupler, these pulses are split
into a strong phase reference, the local oscillator (LO), containing
typically $10^9$~photons per pulse, and a weak signal. The signal pulses are
displaced in the complex plane, with arbitrary amplitude and phase, randomly
chosen from a two-dimensional Gaussian distribution centered at zero and with an
adjustable variance $V_A N_0$. The selected amplitude and phase values are set
by computer-driven electro-optics amplitude and phase modulators placed in the
signal path. Finally, after part of the signal is removed for
synchronization and system characterization purposes (see
Section~\ref{sec:program} for details), Alice's desired modulation variance is
adjusted with a second amplitude modulator and a variable attenuator.

To avoid any polarization and phase drifts that may occur between the signal
and
LO over long-distance transmissions, and thus an incurred additional noise, both
signal and LO pulses need to travel in the same optical fiber. Because of the
simplicity of the corresponding setup, we have opted for time multiplexing,
which is implemented by delaying the LO using an 80~m fiber in its path and then
combining the signal and LO pulses at the output of Alice's setup, as shown in
Fig.~\ref{fig:setup}. Subsequently, the signal and LO pulses, separated by
400~ns, are sent to Bob via the quantum channel, which is a standard single-mode
optical fiber coil.

Bob passively demultiplexes the signal and LO using a 90/10 fiber-optic coupler,
thus introducing a $10\%$ loss in the signal. Then, Bob selects the quadrature
to be measured by adjusting the measurement phase with a computer-driven phase
modulator placed in the LO path. Another 80~m delay line, placed now in the
signal path, results in the signal and LO pulses overlapping at the output
beamsplitter of the interferometer. To ensure a good interference contrast, the
path difference between the signal and LO has to be adjusted to less than a
centimeter. The selected quadrature measurement is then obtained with an
all-fiber shot-noise limited time-resolved pulsed homodyne detection system.
This measurement consists in the substraction of the photocurrents of two fast
InGaAs photodiodes followed by a low noise charge amplifier and a constant gain
amplifying stage.

The choice of the coupling ratios for the multiplexing and demultiplexing
couplers
of the signal and LO in the described setup is the result of a trade-off. First,
the intensity of the LO at the homodyne detection stage needs to be sufficiently
high for the shot noise to be significantly higher than the electronic noise of
the detector. Typically, more than $10^7$~photons per pulse are required for
this purpose. Second, signal losses at Bob's site need to be minimized because
they directly contribute to errors that decrease the mutual information between
Alice and Bob. The coupling ratios quoted in Fig.~\ref{fig:setup} reflect this
trade-off and fulfill the intensity level constraints and the stability
requirements of the system.

\subsection{System automation}
\label{sec:program}

Alice and Bob communicate via a synchronous automatic data processing software,
described in detail in~\cite{lodewyck:pra}. A key transmission is composed of
independent blocks containing $50\,000$ pulses.
Among these pulses, $10\,000$ are used as test pulses which have agreed
amplitude
and phase values, and serve the dual purpose of synchronizing Alice and Bob and
determining the relative phase between the signal and the LO. An additional
random subset of the raw data, typically $5\,000$ pulses, is used for
statistical evaluation of the channel parameters, namely the channel
transmission
$T$ and the excess noise $\varepsilon$, over this subset. In addition, the
signal level sent by Alice and LO level received by Bob are monitored in
real-time on an individual pulse basis. Note that monitoring the LO level for
each pulse
also serves the purpose of avoiding potential
``side-channel" attacks which might tamper classically with the LO
intensity. When
combined with an appropriate calibration, these measurements allow us to obtain
an
accurate estimate of the shot noise level at Bob's site, which is used as a
normalization factor. From this calibration, we can then determine the second
order moments of the data distribution between Alice and Bob: $V_A N_0$, $V_B
N_0$, and
the correlation $\rho$. These moments yield the channel parameters $T$ and
$\varepsilon$, and the information rates. It is important to point out that $T$
is measured both using test pulses of fixed amplitude and a subset of the raw
data, and the agreement between the two values is continuously checked. Taking
into account the fraction of pulses used in each block for synchronization and
system characterization, the repetition rate effectively used for key
distribution is 350~kHz. We note that higher repetition rates up to 1 MHz have
been implemented.

We have designed a software that both manages the interface between Alice and
Bob and ensures	 proper hardware operation, with features aiming towards the
complete automation of the CVQKD system. A software feedforward loop
automatically adjusts every 10 seconds the bias voltages that need to be
applied
to the amplitude modulators in Alice's site, thus compensating for thermal
drifts that occur in the timescale of a few minutes. Furthermore, Alice's
output
modulation variance is stabilized and controlled by a software loop to prevent
natural drifts of the system from modifying the signal to noise ratio (SNR).
This keeps the SNR within the range compatible with the reconciliation codes. At
Bob's site, another software drives Bob's phase generator, using binary
numbers
provided by a quantum random number generator (id Quantique). This chosen phase
is later compensated by the measurement of the relative phase between the signal
and LO. The implementation of these
automated procedures ensures a stable and reliable system operation with minimal
human intervention. Finally, with the exception of the 50/50 coupler at the
input of the homodyne detection system, the setups of Alice and Bob consist
entirely of polarization-maintaining components. This means that polarization
control is only required before the homodyne detector, and to compensate for
polarization drifts in the quantum channel. The use of a
polarization-maintaining homodyne detector and a software-driven low-loss
dynamic polarization controller placed at the input of Bob's setup allows the
implementation of the required compensation while only inducing
reasonable losses to the signal, and leads to fully automatic operation
of the QKD system.

\subsection{Experimental parameters and noise analysis}
\label{sec:noise}

In the previous sections we have described a system that produces correlated
Gaussian-distributed continuous variables at an effective rate of 350~kHz. In
order to obtain the raw key distribution rate from these correlations, we need
to
evaluate the losses and noise sources that are present in the system and
degrade
its performance. At Alice's site, several sources of loss are present in the
signal path, namely modulators (2.5~dB each), polarizers (0.3~dB),
connectors (0.2~dB) or couplers. These losses do not affect the system
performance because the signal level is set at Alice's output. However, the
losses in the LO path need to be controlled so that the intensity
level is
sufficient for the homodyne detection measurement, as we discussed in
Section~\ref{sec:setup}. The quantum channel is a 25~km single-mode optical
fiber, which presents a loss of 5.2~dB. At Bob's site, the losses of the
components in the signal path deteriorate the
transmission signal to
noise ratio (SNR) and thus the amount of key information exchanged between Alice
and Bob. Therefore, these losses must be minimized. To benefit from
the
``realistic mode" assumption described in Section~\ref{sec:rates}, it is
important to carefully calibrate Bob's setup efficiency $\eta$ because
overestimating this value could open a security loophole in the system. The
present overall efficiency, including the homodyne detection efficiency, is
$\eta = 0.606$. Taking into account the measured value $T = 0.302$ for the
channel transmission efficiency, we find that the overall transmission between
Alice and Bob is $\eta T = 0.183$.

In addition to the noise introduced by the channel and homodyne detection
losses, an excess noise due to technical limitations as well as an electronic
noise introduced by the homodyne detection system are present in the system. The
noises contributing to the excess noise $\varepsilon$ can be
independently determined from the experimental data, and lead to an excess noise
of $\varepsilon = 0.005$ shot noise units for a modulation variance $V_A N_0 =
18.5 N_0$. As discussed in Section~\ref{sec:program}, during key transmission
the excess noise is measured by the data processing software. This measurement
was checked experimentally with the implementation of an intercept and resend
attack, where we expect an excess noise of two shot noise units, corresponding
to the ``entanglement breaking" bound for the coherent-state CVQKD
protocol~\cite{lodewyck:prl}. It is important to point out that, in principle,
the excess noise is not caused by Eve and could be considered inaccessible to
her. However, because the diode phase noise and the modulation noises
depend on the
modulation settings, it is difficult to accurately estimate and calibrate this
excess noise. Thus, to avoid compromising the security of our implementation we
assume that it is in fact generated and controlled by Eve. Finally, the homodyne
detector electronic noise contributes $\Velec = 0.041$ shot noise units to the
total noise.

With the help of the equations given in Section~\ref{sec:rates}, the noise
measurements described above lead to the raw secret rates:
\begin{eqnarray}
&I_{AB} = 365\ \textrm{kb/s},\quad I_{BE} = 313\ \textrm{kb/s}&\nonumber\\
&\mathbf{\Delta I^{\text{Shannon}}} = \mathbf{52\ kb/s}& \nonumber\\
&I_{AB} = 365\ \textrm{kb/s},\quad \chi_{BE} = 316\ \textrm{kb/s}&\nonumber\\
&\mathbf{\Delta I^{\text{Holevo}}} = \mathbf{49\ kb/s}&\nonumber
\end{eqnarray}

To obtain a secret key from this information, available in the form of raw
Gaussian correlated data, we have to efficiently extract a string of secret bits
from this data. This is the subject of the next section, which focuses on the
Shannon rate. A very similar procedure can be applied to the Holevo rate.

\section{Reconciliation of continuous Gaussian variables}
\label{sec:reconciliation}

In photon-counting based QKD protocols, data is readily available
as binary digits and can be easily processed for error correction and privacy
amplification using well-known protocols such as Cascade~\cite{cascade} or
Winnow~\cite{winnow}. The amount of secret key that can be extracted from these
error-correction algorithms depends on the error rate of the noisy key. On the
other hand, continuous-variable QKD protocols only provide Alice and Bob with
sequences of correlated Gaussian symbols, from which various noise variances are
determined~\cite{lodewyck:prl}. In particular, the variance of the excess noise
is the analog of the error rate in photon-counting QKD protocols. From
these variances, the mutual informations $I_{AB}$ and $I_{BE}$ can be deduced,
and thus the secret key rate. Therefore, for
CVQKD protocols high secret key distribution rates are attainable, provided that
the secret information $\Delta I^{\text{Shannon}} = I_{AB}-I_{BE}$ available
from the raw Gaussian data can be efficiently extracted.
From a strict
information-theoretic perspective there exists
no fundamental limitations to this extraction process. However, in practice,
error correction requires more information exchange than predicted by Shannon's
theory. The raw secret
information rate is therefore decreased to the effective secret rate $\Delta
I^{\text{Shannon}}_\subs{eff}
= \beta
I_{AB}-I_{BE}$, where the efficiency $\beta<1$ characterizes how close the
reconciliation algorithm operates with respect to the Shannon limit (see
Section~\ref{sec:algo}). Since the maximum achievable transmission distance
ultimately
depends on the value of $\beta$, designing efficient reconciliation algorithms
is one of the challenges of CVQKD. The efficiency of
the first
reconciliation algorithms used for CVQKD~\cite{VanAssche2004,Nguyen2004} did
not reach
80\% for significant line losses, which limited the maximum transmission
distance to less than 20~km. In what follows,
we first briefly review the key principles of a more efficient algorithm
presented in~\cite{Bloch2006}, and then focus on its practical implementation.

\subsection{Multilevel reverse reconciliation with Low-Density
Parity-Check codes}
\label{sec:algo}

Let $X$ denote the random variable representing Alice's Gaussian symbols and $Y$
the one representing Bob's symbols. In theory Alice and Bob should be able to
extract up to $I(X;Y)$ common bits from their correlated sequence. Following the
idea of~\cite{VanAssche2004}, Bob first quantizes his data to obtain discrete
symbols, represented by the variable $\mathcal{Q}(Y)$, and assigns a binary
label to each of them. The quantization necessarily reduces the amount of
extractable information $I(X,\mathcal{Q}(Y)) < I(X;Y)$; however, the penalty can
be made negligible by choosing the quantizer $\mathcal{Q}$ to maximize the
mutual information $I(X;\mathcal{Q}(Y))$. In order to allow Alice to recover his
bit sequence without errors, Bob should then send redundant information, such as
the value of parity-check equations. The theoretical number of such redundancy
bits is $H(\mathcal{Q}(Y)|X)$~\cite{Slepian1973}, however in practice perfect
error correction is only possible when the number of bits disclosed
$M_\subs{rec}$ exceeds this limit. The efficiency $\beta$ of a practical
reconciliation algorithm is then defined as:
\begin{equation}
	\beta = \frac{H(\mathcal{Q}(Y))-M_\subs{rec}}{I(X;Y)}\leq
	\frac{I(X;\mathcal{Q}(Y))}{I(X;Y)}\leq 1.
\end{equation}
\begin{figure*}[tb]
	\includegraphics[width=.8\textwidth]{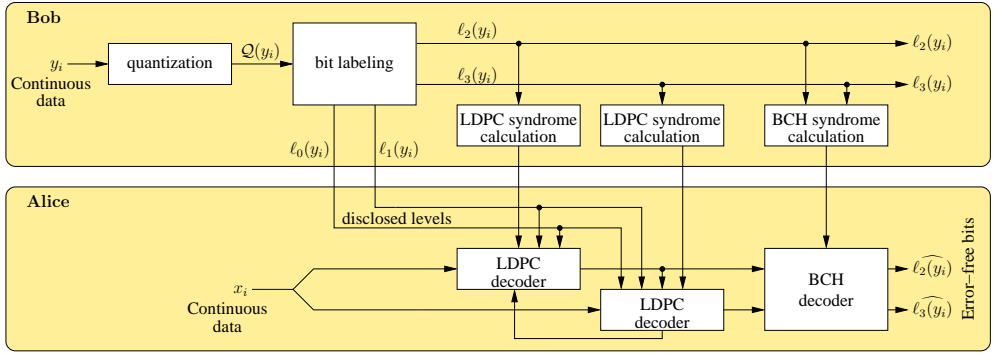}
	\caption{(Color online) Principle of multilevel reconciliation with LDPC
codes.}
	\label{fig:reconciliation}
\end{figure*}
The principle of our reconciliation scheme is shown in
Fig.~\ref{fig:reconciliation}. Once his Gaussian symbols $\left\{y_i\right\}$
have been quantized into $\left\{\mathcal{Q}(y_i)\right\}$, Bob assigns a
$L$-bits binary label $\left\{\ell_j(y_i)\right\}_{j=0..L-1}$ to each of them,
and
calculates a set of parity bits (or \emph{syndromes}) for each individual level
$j$ of label bits. In our case, the number of levels in the multilevel
reconciliation is $L = 4$. This particular encoding incurs no loss of
performance, and
the ideal number of parity bits to disclose at each level can be precisely
calculated~\cite{Bloch2006}. The levels corresponding to the less significant
bits often require almost as many parity bits as there are data bits, and
in this case Bob can simply disclose the entire level. For the levels
corresponding to more
significant bits, the parity bits are calculated according to the parity-check
matrix of Low Density Parity Check (LDPC) codes. Finally, a few extra parity
bits are obtained by applying an algebraic code (such as a BCH
code~\cite{Lin2004}) to the whole data sequence.

Alice retrieves Bob's bit sequence by decoding the bit levels successively,
using her Gaussian symbols $\left\{x_i\right\}$ and the syndromes
sent by
Bob. As illustrated in Fig.~\ref{fig:reconciliation}, the decoding of a level
also exploits the results obtained at the decoding of the previous levels. The
standard decoding algorithm of LDPC codes
(\emph{Sum-Product}~\cite{Richardson2001a}) may sometimes leave a few errors
uncorrected, however the parity bits obtained with the algebraic code are
usually sufficient to correct them.

In comparison with the algorithm proposed in~\cite{VanAssche2004}, which
introduced slice reconciliation with turbo codes, the good efficiency
obtained with this algorithm stems from three key features. First, codes applied
at each level are state-of-the-art LDPC error correcting codes. Then, the
reliability associated to the decision (so-called \emph{soft decoding}) output
from
these codes is used as an \emph{a priori} for the decoding of other levels,
rather than only the bit estimate issued by each decoder. Finally, we allow
several iterations between the levels. In fact, soft decoding enables us to
start the decoding of a level $j$ even if the previous level $j - 1$ has not
been
successfully corrected. A later attempt at decoding level $j - 1$ might benefit
from a partial decoding of level $j$ and could terminate successfully. In
addition, the exchange of information during the whole reconciliation process is
unidirectional, which leaves no ambiguity on the information intercepted by the
eavesdropper.

It was shown in~\cite{Bloch2006}, that LDPC codes with a block
length of $200\,000$ bits were sufficient to achieve efficiencies above 85\%
over a wide range of SNR.

The efficiency $\beta$ characterizes the ultimate performance of a
reconciliation algorithm, however it only assesses its performance from an
information-theoretic standpoint and does not account for the associated
computational complexity. In practice, the latter is of uttermost importance if
one hopes to obtain high secret key distribution rates. Before going on to the
details of the implementation of our algorithm, it is worthwhile discussing the
trade-off between efficiency and decoding complexity. Increasing the
reconciliation efficiency while still maintaining an arbitrarily low probability
of decoding error would require LDPC codes operating closer to the Shannon limit
as well as many more iterations in the decoding process. It is clear that the
code block length and decoding complexity of this scheme would then quickly
become prohibitive. However, a better trade-off can be obtained by maintaining
an arbitrarily low probability of undetected errors. In fact, if the
reconciliation algorithm detects all decoding failures with high probability but
fails to correct errors with probability $p_\subs{fail}$, the effective secret
information rate becomes $\Delta I^{\text{Shannon}}_\subs{eff} = \left(\beta
I_{AB}-I_{BE}\right)\left(1-p_\subs{fail}\right)$. It is difficult to obtain an
analytical expression of $p_\subs{fail}$ as a function of $\beta$ due to the
iterative nature of the decoding process, however we observed via Monte-Carlo
simulation that $\beta$ could be increased by a few percents without too much
sacrifice on $p_\subs{fail}$. Table~\ref{tab:simulations} shows our simulation
results obtained for a mutual information $I(X;Y)=1$ bit/symbol, a 4-bit
quantization, length $200\,000$ LDPC codes, and for a BCH code rate of 0.998 
to obtain the extra parity bits. No undetected errors appeared during the
simulations.

\begin{table}[b]
	\begin{tabular}{|c|c|c|}
		\hline
		LDPC code rates &   $\beta$ &   $p_\subs{fail}$ \\
		\hline
		\hline
		0/0/0.42/0.94 & 86.7\%      &   0   \\
		\hline
		0/0/0.44/0.94 & 88.7\%      &   $10^{-4}$   \\
		\hline
	\end{tabular}
	\caption{Simulation results.}
	\label{tab:simulations}
\end{table}

\subsection{Practical implementation}

As mentioned earlier, the efficiency of the reconciliation strongly depends on
how close the LDPC codes operate with respect to their ideal limit. High
efficiency is therefore only achievable with relatively large block length
(typically over $100\,000$ bits) and randomly constructed
codes~\cite{Richardson2001a}, which makes a hardware implementation of the
algorithm unrealistic. To date, high decoding throughputs on Field Programmable
Gated Arrays (FPGAs) have only been obtained with structured short length codes,
which specific structure allowed a certain amount of parallelism. In our
situation, a software implementation of the algorithm turned out to be the only
viable solution. Typical software implementations of the Sum-Product decoding
algorithm are rather slow, however the execution speed can be substantially
improved by performing fixed-point operations and approximating computationally
intensive functions with table look-ups~\cite{Hu2001}. These simplifications
yield a significant overall speed gain with a negligible performance
degradation. The convergence speed of the LDPC codes can also be accelerated by
using a modified version of the standard Sum-Product decoding
algorithm~\cite{Baynast2005}. A simple change in the scheduling of the decoding
reduces the number of iterations by a factor almost two without any penalty in
terms of performance.

In the situation of interest for CVQKD, most of the complexity of the
reconciliation algorithm comes from the use of two LDPC codes of same block
length. The decoding complexity depends on many parameters, such as the number
of iterations performed during the decoding of each code, the number of times
each level is decoded, the average number of terms involved in parity-check
equations, etc. For a desired level of performance, there exists no generic
method for finding a set of parameters minimizing the complexity because all
parameters interplay in a complex manner. For instance, choosing ``better''
codes operating closer to the Shannon limit could reduce the number of
iterations required in each LDPC decoder, but the size of the parity-check
equations would generally increase. Likewise, increasing the number of
iterations within a LDPC decoder may sometimes reduce the number of iterations
required between different decoders. Hence the choice of the parameters
described hereafter results from many heuristic optimizations.

\subsection{Optimal reconciliation parameters}

\begin{figure}[tb]
	\includegraphics[width=.45 \textwidth]{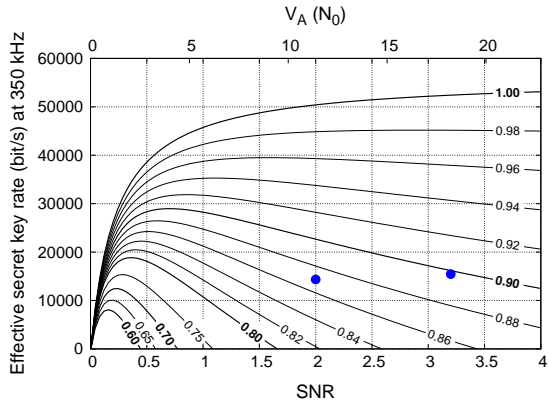}
	\caption{(Color online) Effective key distribution rate as a function of
the SNR, or
equivalently, as a function of the modulation variance $V_A N_0$. We assume
a linear dependence of $\varepsilon$ on $V_A$, and use the experimental
parameters
given in Section~\ref{sec:implementation}. The curves show the key distribution
rate for values of $\beta$ ranging from 0.6 to 1.0, and the filled circles show
the efficiency we actually achieve for different SNR.}
	\label{fig:VA}
\end{figure}

Although code parameters cannot be deduced from an analytical calculation, the
optimal modulation variance is the result of a quantitative compromise. The
reconciliation efficiency only depends on the SNR of the transmission, which,
for a given distance, is an increasing function of the modulation variance
$V_A N_0$. However, as shown in Fig.~\ref{fig:VA}, the effective secret rate 
$\Delta I^{\text{Shannon}}_\subs{eff} = \beta I_{AB}-I_{BE}$ as a function of
$V_A$ and SNR clearly exhibits an
optimal value for $V_A$. For the targeted transmission distance of
25~km
($T=0.302$) and considering the detector efficiency and noise discussed in the
previous
section, which require a reconciliation efficiency above 80\%, we obtained
the best performance/complexity trade-off with the following parameters:
\begin{itemize}
    \item modulation variance $V_A N_0=18.5 N_0$ ($I(X;Y)=1.045$ bit/symbol),
    \item quantization with 16 equally spaced intervals
($I(X;\mathcal{Q}(Y))= 1.019$ bit/symbol), ideally requiring 4 codes
with rates 0.002/0.013/0.456/0.981,
    \item practical codes rates 0/0/0.42/0.95, yielding an efficiency
$\beta=0.898$.
\end{itemize}
These reconciliation parameters are adjusted as the line parameters (namely the
excess noise) fluctuate, and yield the following secret key distribution rates:
\begin{eqnarray*}
\mathbf{\Delta I^{\text{Shannon}}_{\text{eff}}}  =  \mathbf{15.2\ kb/s}
\qquad
\mathbf{\Delta I^{\text{Holevo}}_{\text{eff}}}  =  \mathbf{12.3\ kb/s}
\end{eqnarray*}

Since the LDPC codes are very demanding in computing power, the reconciliation
speed is directly affected by the processor speed. The use of one core of a
dedicated Core 2 Duo Intel processor leads to a reconciliation speed of
$40\,000$
Gaussian symbols/s, while using a NVidia GTX 7950 graphics processor allows a
speed of
$63\,000$ symbols/s, to be compared with the current repetition rate of
$350\,000$ symbols/s. Taking into account this speed limitation, the final (net)
secure key distribution rates are:
\begin{eqnarray}
	&\textrm{Using a Core 2 Duo CPU:}&\nonumber\\
	&\mathbf{\Delta I^{\text{Shannon}}_{\text{net}}}  = \mathbf{1.7\ kb/s}
\qquad\mathbf{\Delta I^{\text{Holevo}}_{\text{net}}}  =  \mathbf{1.4\
kb/s}&\nonumber\\
	&\textrm{Using a GTX 7950 GPU:}&\nonumber\\
	&\mathbf{\Delta I^{\text{Shannon}}_{\text{net}}}  = \mathbf{2.7\ kb/s}
\qquad\mathbf{\Delta I^{\text{Holevo}}_{\text{net}}}  =  \mathbf{2.2\
kb/s}\nonumber&
\end{eqnarray}

We note that the reconciliation procedure described above has been optimized for
the case of the Shannon entropy, and further optimization should be considered
to achieve a higher Holevo rate.

\section{Privacy amplification}

At the end of the reconciliation process, the classical error correction
algorithm outputs blocks of $b = 400\,000$~bits (\emph{i.e} the two
most significant quantization levels of blocks of $n=200\,000$ continuous
variables), and each of them needs to be compressed into a much shorter secret
key
of typically $k = 10\,000$ secret bits, depending on the measured secret key
distribution rate. In order not to affect the overall
classical processing
speed, this particularly large input size requires us to use fast privacy
amplification algorithms. Privacy amplification~\cite{bennett:pa} consists in
randomly choosing a
\emph{hash function} mapping bit strings of length $b$ to bit strings of length
$k$, among a suitable set of these functions called a \emph{family}. The
probability of success of these algorithms is characterized by the universality
$\epsilon$ of the family of hash functions, and the security parameter $s$,
\emph{i.e.} the number of bits that are sacrificed during the amplification
process. Quantitatively, the probability that Eve knows one bit of the final key
is about $\max(2^{-s},\epsilon - 1)$~\cite{gilles:book}. For universal families
of hash functions, defined by $\epsilon = 1$, only the security parameter $s$ is
therefore relevant. The size of the resulting secret key is then $k =
n\Delta I^{\text{Shannon}}_\subs{eff} - s$.

The simplest practical universal family of hash functions is the
multiplication by a random element of the Galois field $GF(2^l)$ with
$l>b$~\cite{bennett:pa}. The
number theoretic transform (NTT), a FFT-like algorithm in $GF(2^l)$ enables us
to rapidly perform this multiplication~\cite{gilles:book}. Still, the
amplification of $400\,000$
bits with this algorithm takes about 10 seconds on an average desktop computer,
which is about as long as the whole reconciliation process, thus significantly
decreasing the final secret key rate.

To avoid this long computation time, we use instead a non-universal family of
hash functions based on the NTT described in~\cite{gilles:book} (section 7.3.3).
In this algorithm, we first convert the incoming bit string into a vector of
$L_p$
elements of the Galois field $GF(p)$ ($L_p = 2^{14}$ and $p = 33\,832\,961$ are
suitable for our input string length). Then we compute the inverse NTT of the
component-wise product of the generated vector with a random vector with no
zero element. The hash output is then obtained by converting back the result to
a bit string, which is then truncated to the desired key length. This hash
function evaluation only requires a few tens of milliseconds, but its
universality is $\epsilon_1 = 1 + \frac k p \simeq 1 + 5\cdot 10^{-4}$, allowing
for security parameters up to only about 10. To overcome this problem, we
combine this algorithm with the universal
($\epsilon_2 = 1$) family of hash functions based on the
multiplication in $GF(2^m)$. For this, we first non-universally hash our $b$
bits into $m = 19\,937$~bits for which we know a Galois field, and then
universally hash these resulting bits into $k \simeq 10\,000 $~bits. Although
this second hashing algorithm is much slower, the execution time is still
tolerable due to the reduced input size. The universality of the total composite
hashing is $\epsilon_c = 2^{k -19\,937}\epsilon_1 + \epsilon_2$~\cite{stinson},
and so $\epsilon_c -1$ is small enough to allow virtually any reasonable
security parameter. On a desktop computer, the total hashing time is $0.27$~s
per block, of which $0.2$~s are consumed by the second hashing.

\section{Generation of a secret key over a 25 km long fiber}

\begin{figure}[tbh]
	\includegraphics[width=.45\textwidth]{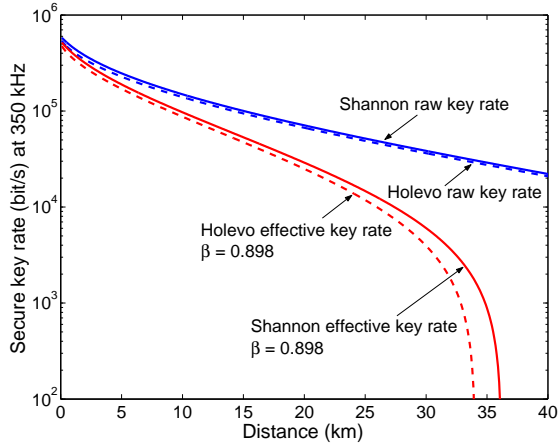}
	\caption{(Color online) Raw and effective key distribution rates for
communication
secure against individual (Shannon) and
collective (Holevo) attacks. The reconciliation efficiency for the effective key
distribution rates is assumed to be $\beta = 0.898$. The parameters used in the
calculations are $V_A N_0 = 18.5 N_0$, $\varepsilon = 0.005$, $\eta = 0.606$,
$\Velec = 0.041$, the effective repetition rate is 350~kHz, and the fiber
loss is assumed to be 0.2~dB/km.}
	\label{fig:rates}
\end{figure}

To finalize our CVQKD system, we designed a software
implementing
the classical channel between Alice and Bob. This software continuously
retrieves Gaussian data from the software driving the experiment,
and performs error correction and privacy amplification. It features an
authentication backend interface that is currently using the authentication
algorithms developed by the European Integrated Project SECOQC~\cite{secoqc}.
With the system described in the previous sections, which combines CVQKD
hardware and key distillation software, we have been able to transmit a binary
secret key over a 25~km long fiber coil with a final secret key distribution
rate of 2~kb/s. This rate takes into account the entire key distillation
procedure, including the classical channel latency. By evaluating our
transmission parameters for different channel transmissions we obtain the
raw and
effective key distribution rate curves shown in Fig.~\ref{fig:rates}.

\section{Conclusion}

In conclusion, we have presented the implementation of a complete
continuous-variable quantum key distribution system, generating secret keys at
a rate of 
more than 2
kb/s over 25 km of optical fiber. The system is secure against individual and
collective attacks, when using Shannon or Holevo information bounds,
respectively. A single program drives hardware automation, signal modulation and
measurement, and performs authentication, reverse reconciliation, and privacy
amplification. Our QKD setup is therefore fully functional
and meets all aspects required for a field implementation.

Currently, the secret key rate is limited by data processing and data
acquisition, rather than by optical components. Further improvements of the
reconciliation algorithms, as well as the use of faster components (CPUs and
data acquisition cards), should thus lead to a direct enhancement of the key
rate.

\begin{acknowledgments}
We acknowledge contributions of C\'ecile Neu to the initial versions of the
communication software, and improvements by Andr\'e Villing to the
system electronics, especially concerning the homodyne detector. We
also acknowledge the support from the European Union under the projects SECOQC
(IST-2002-506813), COVAQIAL (FP6-511004), and QAP (IST-2001-37559), and from the
IUAP program of Belgian federal government. E.D. acknowledges support from the
European Union through a Marie-Curie fellowship (MEIF-CT-2006-039719) and a
Marie-Curie reintegration grant. R.G.-P. acknowledges the support from the
Belgian foundation FRIA. E.K. acknowledges support of the Brussels-Capital
Region within the program ``Prospective research for Brussels 2006''.
\end{acknowledgments}


\end{document}